\documentclass{aip-cp-aac2016}

\usepackage[numbers]{natbib}
\usepackage{rotating}
\usepackage{graphicx}

\usepackage[utf8]{inputenc}
\usepackage[T1]{fontenc}
\usepackage{lmodern}

\begin{document}

\title{Integrable RCS as a proposed replacement for Fermilab Booster}

\author[aff1]{Jeffrey Eldred\corref{cor1}}
\author[aff1]{Alexander Valishev}

\affil[aff1]{Fermi National Accelerator Laboratory, Batavia, IL, USA}
\corresp[cor1]{Corresponding author: jseldred@fnal.gov}

\maketitle

\begin{abstract}
Integrable optics is an innovation in particle accelerator design that potentially enables a greater betatron tune spread and damps collective instabilities. An integrable rapid-cycling synchrotron (RCS) would be an effective replacement for the Fermilab Booster, as part of a plan to reach multi-MW beam power at 120 GeV for the Fermilab high-energy neutrino program. We provide an example integrable lattice with features of a modern RCS - dispersion-free drifts, low momentum compaction factor, superperiodicity, chromaticity correction, bounded beta functions, and separate-function magnets.
\end{abstract}

\section{INTRODUCTION}

Integrable optics is a development in particle accelerator technology that enables strong nonlinear focusing without generating parametric resonances~\cite{Danilov}. A promising application of integrable optics is in high-intensity rings, where it is necessary to avoid resonances associated with a large betatron tune-spread, while simultaneously suppressing collective instabilities with Landau damping. The efficacy of an accelerator design incorporating integrable optics will undergo comprehensive experimental tests at the Fermilab Integrable Optics Test Accelerator (IOTA)~\cite{Valishev} and the University of Maryland Electron Ring (UMER)~\cite{Ruisard} over the next several years. In this paper, we discuss a potential Fermilab integrable rapid-cycling synchotron (iRCS) as a high-intensity replacement for the Fermilab Booster.

At Fermilab, a core research priority is to improve the proton beam power for the flagship high-energy neutrino program~\cite{Prebys}. In the current running configuration, a 700 kW, 120 Gev proton beam is delivered to a carbon target for the NuMI beamline that supports the NOvA, MINERvA, and MINOS neutrino experiments. Next, the Proton Improvement Plan II (PIP-II) will replace the 400 MeV linac with a new 800 MeV linac that will increase the 120 GeV proton power of the Fermilab complex to 1.2 MW~\cite{PIP2}. 

The next flagship neutrino experiment at Fermilab will be the LBNF/DUNE~\cite{DUNE}. The P5 Report referred to LBNF as ``the highest priority project in its lifetime,'' and set a benchmark for a 3$\sigma$ measurement of the CP-violating phase over 75\% the range of its possible values~\cite{P5}. The P5 benchmark for the CP-violating phase corresponds to a 900 kt$\cdot$MW$\cdot$year neutrino exposure requirement~\cite{DUNE,Prebys}. For a proton power of 1.2 MW and a 50 kt LAr detector, 15 years are required to meet that benchmark. For a proton power of 3.6 MW and a 36 kt LAr detector, 7 years are required. 

In order to achieve a 120 GeV proton power significantly beyond the 1.2 MW delivered by PIP-II, it will be necessary to replace the Fermilab Booster with a modern RCS~\cite{Prebys}. The Fermilab Booster is over 45 years old and faces limitations from its magnets and its RF alike. There is no beampipe inside the dipoles, and the magnet laminations generate an impedance instability. The impedance instability provides $\sim$200 kV deceleration during transition crossing at current beam intensities~\cite{Lebedev}. The Booster dipoles are combined function magnets, which constrain tunability and amplify electron cloud instabilities~\cite{Antipov}. The Booster RF cavities underwent a refurbishment process and cooling upgrade in order to achieve a 15-Hz Booster ramp rate~\cite{Pellico}, but the ramp rate will not be able to exceed 20 Hz without replacing the Booster RF entirely.

The Fermilab Booster intensity is also limited by betatron resonance losses incurred by a large Laslett betatron tune-spread. The Laslett tune-spread is given by
\begin{equation}
\Delta \nu (z) \approx \frac{N r_{0}}{2 \pi \epsilon_{N} \beta \gamma^{2}} \left(\frac{\lambda(z)}{\langle \lambda \rangle_{z}} \right) F,
\end{equation}
 where $N$ is the number of particles, $r_{0}$ is the classical electron radius, $\epsilon_{N}$ is the normalized transverse emittance, $\lambda(z)$ is the local charge density at position $z$, and $F$ is a transverse form factor~\cite{Prebys}.

An integrable RCS could manage a high Laslett tune-spread without a costly increase in aperture~\cite{Tang}. Detailed simulations of space-charge dominated beams in integrable lattices is an ongoing work~\cite{Valishev,Bruhwiler}, and the ultimate limitation in Laslett tune-spread is not fully determined.

Table~\ref{Power} shows how 120-GeV beam power corresponds to the Laslett tune-spread the iRCS must support. For this exercise, we consider an iRCS with the same aperture and circumference as the Fermilab Booster, that the iRCS has a 20-Hz ramp rate, and that there is no slip-stacking in the Fermilab Recycler or Main Injector.

\begin{table}[htp]
\begin{centering}
\begin{tabular}{| c | c | c | c|}
\hline
Laslett Tune-spread & RCS & RCS Extracted & Main Injector  \\
Requirement & Intensity & Power at 8 GeV & Power at 120 GeV \\
\hline
-0.27 & 16 e12 & 0.41 MW & 1.2 MW \\
-0.53 & 32 e12 & 0.82 MW & 2.3 MW \\
-0.80 & 48 e12 & 1.20 MW & 3.6 MW \\
\hline
\end{tabular}
\caption{Beam power requirements for iRCS baseline scenario.}
\label{Power}
\end{centering}
\end{table}

Slip-stacking is an accumulation process that doubles the intensity of the Fermilab Main Injector, but the feasibility of slip-stacking beyond the PIP-II era is still under investigation~\cite{Brown,Eldred}. If 20-Hz slip-stacking can be sustained in the Main Injector or Recycler, the 120-GeV beam power would increase by a factor of 1.5 or 2, respectively.

\begin{figure}[htp]
\begin{centering}
\includegraphics[height=180pt, width=270pt]{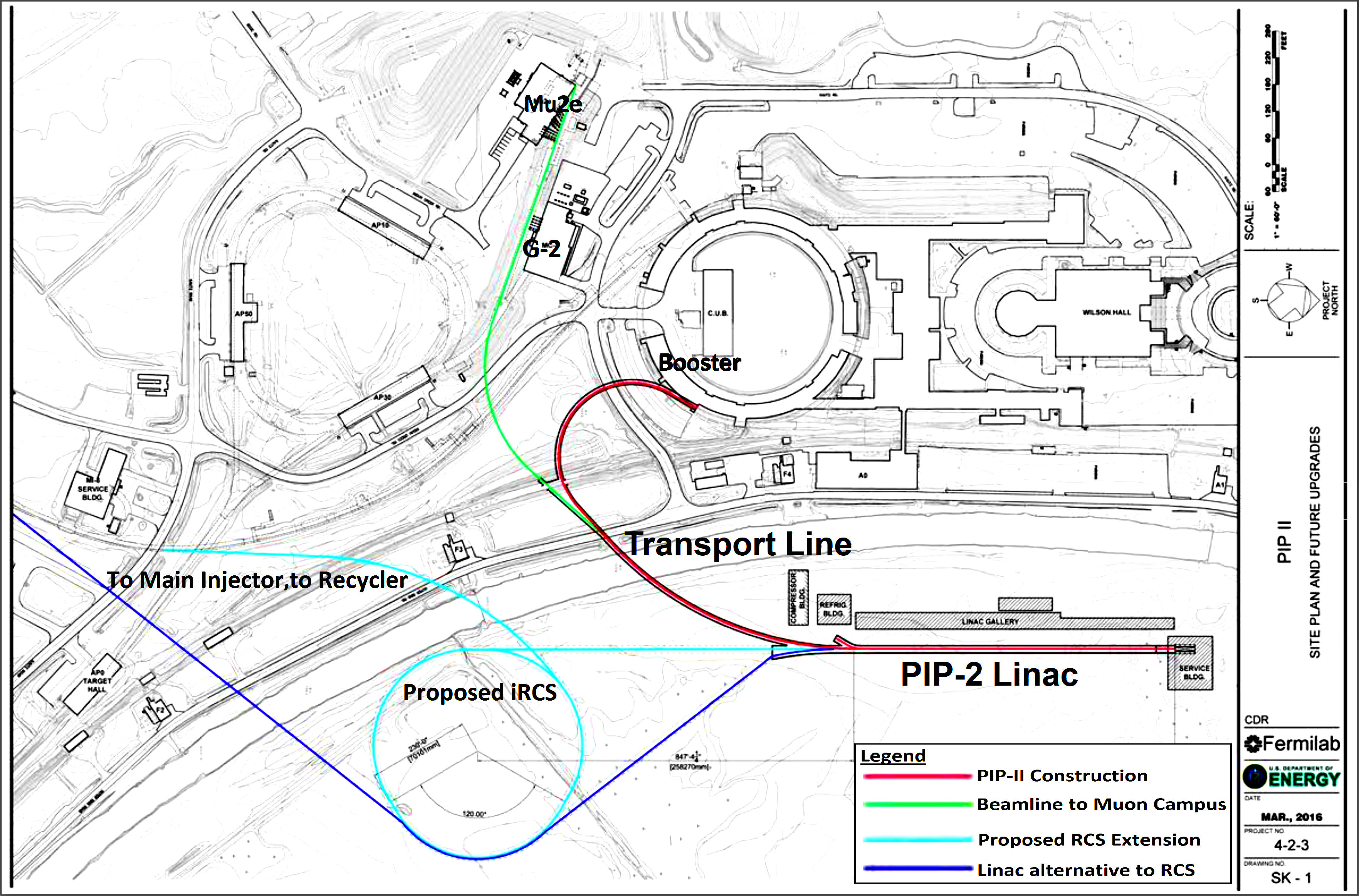}
  \caption{Site location for the proposed iRCS, relative to the PIP-II linac, muon campus, and Main Injector~\cite{Dixon}.}
  \label{Siting}
\end{centering}
\end{figure}

Figure~\ref{Siting} shows one of several siting options for an iRCS and it does not limit the circumference. Excess circumference can raise the costs of the RF system by increasing the revolution frequency~\cite{Tang}. The iRCS ramp rate indirectly affects the 120 GeV beam power by reducing the Main Injector fill time. For a 60-Hz ramp rate, 120 GeV beam power increases by 15\%.

\section{iRCS EXAMPLE LATTICE}

The iRCS should incorporate the innovations in RCS design that have been developed after the Fermilab Booster~\cite{Tang}. Periodicity and bounded beta functions increase the dynamic aperture. Transition crossing can be avoided by designing the lattice with a low momentum compaction factor. Modern RCS design also uses separate-function dipole magnets and long dispersion-free drifts. In this Section, we show an example integrable lattice with these features.

An accelerator can achieve integrable optics with alternating sections of T-inserts and nonlinear magnets~\cite{Danilov}. The T-inserts are arc sections with $\pi$-integer betatron phase-advance in the horizontal and vertical plane. The lattice should be dispersion-free in the nonlinear section, and the horizontal and vertical beta functions should be matched. A special nonlinear elliptical magnet is matched to the beta functions to provide the nonlinear focusing. 

\begin{figure}[h]
\begin{centering}
  \includegraphics[height=200pt, width=360pt]{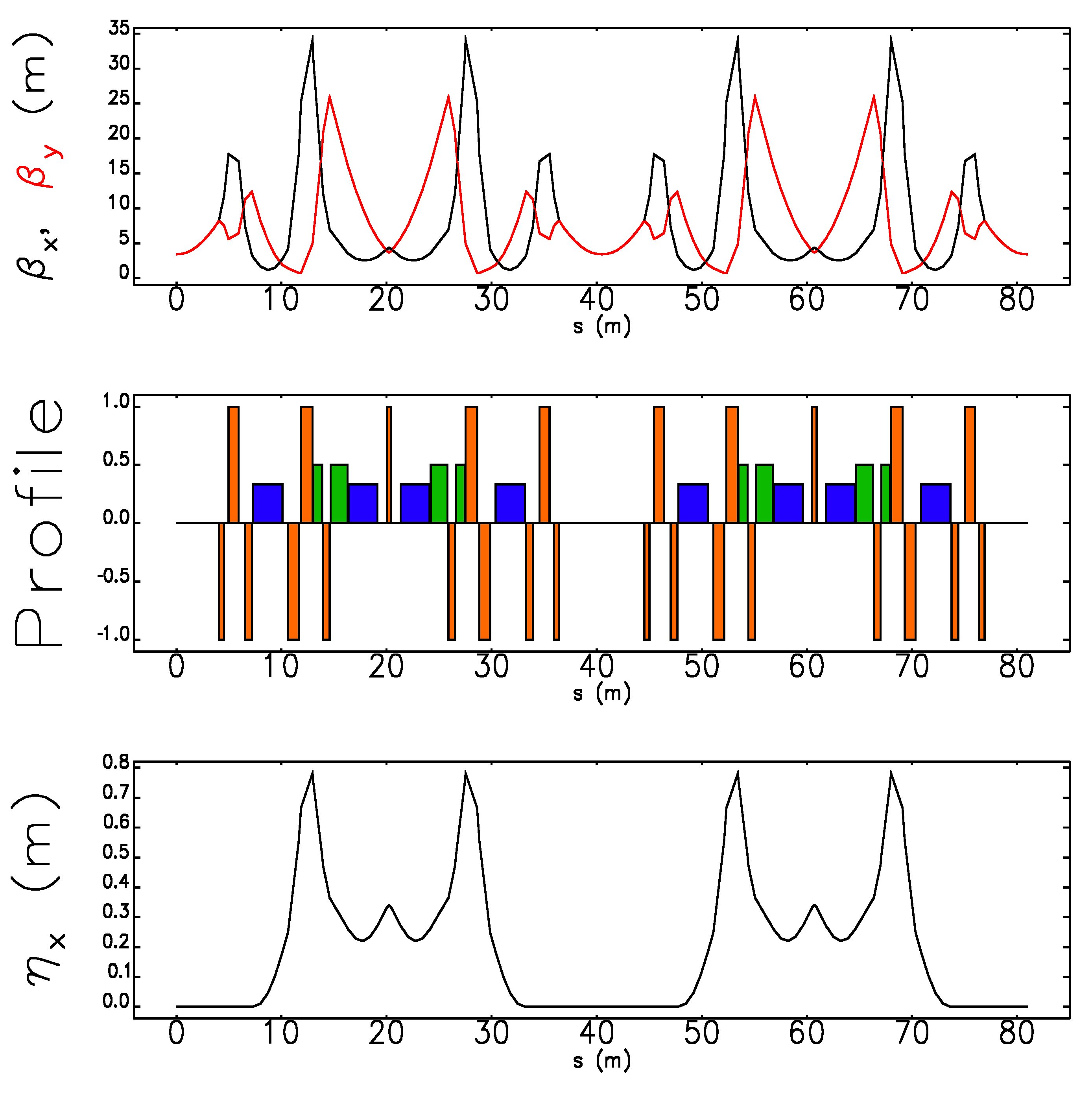}
  \caption{Twiss parameters for one of the six periodic cells. (top) Horizontal and vertical beta functions shown in black and red, respectively. (middle) Location and length of magnetic lattice, elements where dipoles are shown as short blue rectangles, quadrupoles as tall orange rectangles, and sextupoles as green rectangles. (bottom) Linear dispersion function.}
  \label{Lattice}
\end{centering}
\end{figure}

Figure~\ref{Lattice} shows an example iRCS lattice and Table~\ref{Param} shows the key parameters of this lattice. The lattice is composed of 12 identical achromatic arcs and dispersion-free drifts. Every other drift hosts a nonlinear insert, so the lattice forms 6 periodic cells with a T-insert section and a nonlinear insert section. The drifts in the center of each T-insert arc are used for injection, extraction, and RF acceleration. 

\begin{table}[htp]
\begin{centering}
\begin{tabular}{| l | c |}
\hline
Parameter & Value \\
\hline
Circumference & 486 m \\
Periodicity & 6 (12) \\
Vertical Aperture & 5 cm\\
\hline
Maximum Energy & 8 GeV \\
Bend Radius & 15.6 m \\
Peak Dipole Field & 1.25 T \\ 
Peak Quadrupole Field & 25 T/m \\ 
Peak Sextupole Field & 180 T/m$^{2}$ \\ 
\hline
Max Beta Function  & 35 m \\
Max Dispersion & 0.8 m \\
\hline
\end{tabular}
\qquad
\begin{tabular}{| l | c |}
\hline
Parameter & Value \\
\hline
Insertion Length / Cell & 8.1 m \\
Total Insertion Length & 97 m \\
Single Dipole Length & 2.8 m \\
Number of Dipoles & 48 \\
Number of Quadrupoles & 156 \\ 
Number of Sextupoles & 48 \\ 
\hline
Momentum Compaction & 2$ \times 10^{-3}$ \\
Extraction Phase-Slip Factor & -6$ \times 10^{-3}$ \\
Betatron Tunes & 19.7 \\
Linear Chromaticities & -10 \\
Second-order Chromaticities & 50 \\
\hline
\end{tabular}
\caption{Parameters of iRCS Lattice}
\label{Param}
\end{centering}
\end{table}

By extending the design circumference of the iRCS, the design length of the dispersion-free drifts can readily be extended. Here the iRCS accommodates 40 meters of accelerating RF cavities with a 1.2 MV total voltage and a 20-Hz ramp rate. The peak dipole field of 1.25 T was chosen to minimize eddy-current losses. Our design efforts favor perpendicular-biased ferrite RF cavities, which may deliver twice the accelerating gradient of parallel-biased ferrite RF cavities. Perpendicular-biased ferrite cavities are subject of an active R\&D effort at Fermilab~\cite{Romanov}.

The effect of linear chromaticity on integrable motion was examined in \cite{Webb}, and the effect of nonlinear chromaticity was examined in \cite{Cook}. Chromaticity is compatible with integrability if the horizontal and vertical chromaticities are matched. In the example iRCS lattice, the chromaticities were matched and reduced in both the first and second order. The chromaticity values shown in Fig.~\ref{Param} are the result of correction by the sextupole magnets (green in Fig.~\ref{Lattice}).

To preserve integrability, sextupole magnets should also be located so that their effect cancels harmonically (separated by a $\pi$-odd phase-advance)~\cite{Webb}. To maintain the flexibility of the early design, this constraint was not imposed on the example lattice shown here. This constraint can be met by requiring a $\pi$-odd phase-advance for the 12 linear-periodic cells, or by combining into 6 complex linear-periodic cells.

\section{LONGITUDINAL CAPTURE}

The PIP-II linac is designed for a 2 mA beam current with a pulse duration of 0.6 ms to fill the Fermilab Booster to an intensity of $6.4~\times~10^{12}$ protons~\cite{PIP2}. To limit the injection time into the iRCS to 2-3ms, a 5 mA upgrade of the PIP-II linac is necessary.

For a preliminary analysis of longitudinal capture, we considered two scenarios to fill the RCS while controlling longitudinal emittance. In the aggressive chopping scenario, the beam is chopped 50\% at 44 MHz to match the RF buckets. The fill time is 3.1 ms for $48~\times~10^{12}$ protons. In the adiabatic capture scenario, the beam is injected into 0.12 MV RF buckets, and the voltage is raised to 1.2 MV over 0.6 ms. Only 5\% of the beam is chopped, and consequently, the fill time is 1.7 ms for $48~\times~10^{12}$ protons. 

\begin{figure}[htp]
\begin{centering}
\includegraphics[height=200pt, width=360pt]{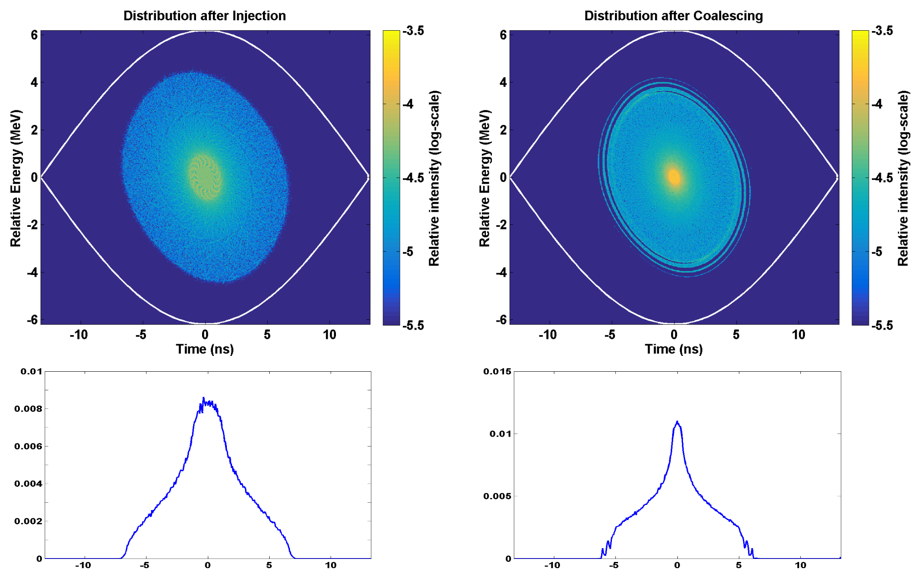}
  \caption{(left) Longitudinal phase-space distribution (top) and longitudinal profile (bottom) for aggressive chopping scenario. (tight) Longitudinal phase-space distribution (top) and longitudinal profile (bottom) for adiabatic capture scenario.}
  \label{Long}
\end{centering}
\end{figure}

Figure~\ref{Long} shows the longitudinal distribution for the aggressive chopping and adiabatic capture scenarios. To achieve a more uniform charge density, the injection energy was painted across a slowly accelerating RF bucket. The injected beam was painted over 30\% of the bucket height, for the aggressive chopping scenario, and 17\% for the adiabatic capture scenario. The acceleration rate was adjusted until the 97\% longitudinal rms emittance of 0.08 eV$\cdot$s was reached, consistent with the PIP-II Booster emittance~\cite{PIP2}. If beam from the iRCS is not slip-stacked, the longitudinal emittance can be a factor of 5 larger before encountering a particle loss limit caused by transition crossing in the Main Injector. If a second harmonic RF cavity is used during injection, a more uniform charge density can be achieved while maintaining this longitudinal emittance~\cite{JPARC,Pham}.

To minimize power losses, the lattice magnets will follow a sinusoidal ramp curve~\cite{Tang}. A corrector circuit will flatten the sinusoidal ramp curve during injection, so that the injection energy consistently matches the lattice energy. For a 20-Hz ramp rate, a 3 ms injection time corresponds to a 6\% of the cycle time and a 2\% amplitude correction.

\section{SUMMARY \& FUTURE WORK}

To achieve multi-MW beam power for the Fermilab high-energy program, an integrable RCS replacement for the Fermilab Booster is an option that merits careful scrutiny. In this paper we explore some of the preliminary design concerns, including integrable RCS lattice design and longitudinal capture. Experimental and numerical work on the interaction between integrable optics and space-charge dominated beams is still ongoing~\cite{Valishev,Ruisard,Bruhwiler}.

Upcoming lattice designs will integrate sextupoles with harmonic cancellation. Thresholds for collective instabilities in the highly nonlinear beam will be analyzed for their impact on the iRCS parameters. The design of the high-power H$^{-}$ stripping foil and injection chicane will also be developed. A second harmonic RF will be added to the longitudinal capture analysis to demonstrate longitudinal uniformity.

\nocite{*}
\bibliographystyle{aipnum-cp}%
\bibliography{bib.tex}%

\end{document}